\documentclass[journal,9pt]{IEEEtran}

\usepackage{newcent}

\usepackage[cmex10]{amsmath}
\usepackage{amssymb}

\usepackage[pdftex]{graphicx}

\newcommand{\Ba}{B_2}
\newcommand{\Gx}{G_x}
\newcommand{\Gy}{G_y}
\newcommand{\Gxptilde}{G_{1}^{x}}
\newcommand{\Gyptilde}{G_{1}^{y}}
\newcommand{\Ltot}{L}
\newcommand{\conn}{\mathbb{C}}
\newcommand{\link}{\mathbb{L}}
\newcommand{\wdm}{\mathbb{W}}

\newcommand{\Gsig}{G}
\newcommand{\Gnli}{G^\text{NLI}}
\newcommand{\Gase}{G^\text{ASE}}
\newcommand{\Gsci}{G^\text{SCI}_m}
\newcommand{\Gxci}{G^\text{XCI}_{mm'}}

\newcommand{\dilog}{\operatorname{Li_2}}

\newcommand{\fa}{\nu}
\newcommand{\fb}{\nu'}

\newcommand{\overlay}[3]{\makebox[0mm][l]{\hspace*{#1}\raisebox{#2}[0ex][0ex]{#3}}}

\begin{document}
\title{Modeling of Nonlinear Signal Distortion\\in Fiber-Optical Networks}

\author{Pontus Johannisson and Erik Agrell% <-this % stops a space
\thanks{Pontus Johannisson is with the Photonics Laboratory, Department of
        Microtechnology and Nanoscience, Chalmers University of
        Technology, SE-412\,96 Gothenburg, Sweden.}% <-this % stops a space
\thanks{Erik Agrell is with the Communication Systems Group, Department of
        Signals and Systems, Chalmers University of Technology,
        SE-412\,96 Gothenburg, Sweden.}}% <-this % stops a space

\markboth{}%
{}

\maketitle

\begin{abstract}
%\boldmath
A low-complexity model for signal quality prediction in a nonlinear fiber-optical network is developed. The model, which builds on the Gaussian noise model, takes into account the signal degradation caused by a combination of chromatic dispersion, nonlinear signal distortion, and amplifier noise. The center frequencies, bandwidths, and transmit powers can be chosen independently for each channel, which makes the model suitable for analysis and optimization of resource allocation, routing, and scheduling in large-scale optical networks applying flexible-grid wavelength-division multiplexing.
\end{abstract}

\begin{IEEEkeywords}
Optical fiber communication, Optical fiber networks, Fiber nonlinear optics,
Wavelength division multiplexing.
\end{IEEEkeywords}

\IEEEpeerreviewmaketitle

\section{Introduction}
\IEEEPARstart{O}{ptical} long-haul communication research has during the latest decade become completely focused on coherent transmission. The data throughput has been increased by better utilization of the available bandwidth and receiver digital signal processing (DSP) has allowed many signal impairments to be compensated for. Further significant increase of the data rate is expected through the use of multicore/multimode fibers~\cite{essiambre_2012_procieee}, but while this development is exciting, there are also significant challenges. For example, these approaches call for the deployment of new fibers.

It is also important to use available resources to the greatest possible extent. The increase in spectral efficiency enabled by coherent communication is actually an example of this. For example, using a channel spacing of 50~GHz, it is now possible to transmit 100~Gbit/s using polarization-multiplexed quadrature phase-shift keying. Comparing this with the traditional 10~Gbit/s using on--off keying, the data throughput is increased by a factor of ten. In this work, the focus is on optical networking. By developing routing algorithms with awareness of the nonlinear physical properties of the channel, it would be possible to operate optical links closer to the optimum performance, and this would further increase the throughput in optical communication networks.

Optical networks are not as flexible as their electronic counterparts, but there are several efforts that aim at improving the situation by investigating, \textit{e.g.}, cognitive~\cite{miguel_2013_ofc} and elastic~\cite{rival_2011_ofc} optical networks. An increased flexibility requires, \textit{e.g.}, hardware routing of channels in the optical domain and a monitoring system responsible for the scheduling of the data streams. It is also desirable that the coherent transmission and detection can be done using a number of different modulation formats and symbol rates, chosen dynamically in response to time-varying traffic demands and network load. Hence, the traditional wavelength-division multiplexing (WDM) paradigm, in which the available spectrum is divided into a fixed grid of equal-bandwidth channels, is being replaced by the concept of \emph{flexible-grid} WDM~\cite{gerstel_2012_cm}.

The scheduling algorithm requires a nonlinear model of the physical layer. While the problem of linear routing and wavelength assignment is well investigated, see, \textit{e.g.}, \cite{ramaswami_1995_ton}, no nonlinear model that combines reasonable accuracy and low computational complexity seems to have been published. Both these properties are necessary in order to be able to find a close-to-optimal solution in real time. This excludes simulations of the nonlinear Schr\"odinger equation as an alternative, but the recently suggested Gaussian noise (GN) model~\cite{poggiolini_2011_ptl2,carena_2012_jlt,poggiolini_2012_jlt} provides a tool to approach this question. Unfortunately, also the general formulation of the GN model is computationally complex and further simplification is necessary. In this paper, we start from a description of how a model suitable for network optimization can be formulated and derive such a model from the GN model.

The organization of this paper is as follows. In Sec.~\ref{sec_approach}, the considered problem is stated and the approach is outlined. The main assumptions and approximations are given in Sec.~\ref{sec_model}. The model for a multichannel fiber span is derived in Sec.~\ref{sec:span}, and it is expanded into a network model in Sec.~\ref{sec:network-model}. After a brief discussion about the validity of the model assumptions, the paper is concluded in Sec.~\ref{sec_discussion}.

\section{Problem Statement}
\label{sec_approach}

\subsection{Network Topology and Terminology}
A network consists of a number of \emph{nodes,} \textit{i.e.}, transceivers or routing hardware components, connected by optical communication \emph{links}. Each link consists of $N$ concatenated \emph{fiber spans,} which each consists of an optical fiber followed by an erbium-doped fiber amplifier (EDFA). Each link can transmit $M$ simultaneous \emph{channels} using WDM.

In such a network, a large number of \emph{connections} between transceiver nodes are established. For each connection there is a \emph{route,} \textit{i.e.}, a set of fiber links that connect the transmitting and receiving nodes via a number of intermediate nodes. We consider an all-optical network, implying that the signal is in the optical domain throughout the path. The intermediate nodes typically consist of reconfigurable add/drop multiplexers and may also include wavelength conversion.

\subsection{Signal Degrading Mechanisms}
For any connection through the network, there will be signal degradation caused by a number of mechanisms. The two most fundamental ones are amplifier noise and nonlinear signal distortion due to the Kerr nonlinearity. While the amplified spontaneous emission from optical amplifiers is easy to model, the latter presents a big challenge. Additional degrading effects include the finite signal-to-noise ratio (SNR) already at the transmitter, which may be important for large quadrature amplitude modulation constellations, and the crosstalk between different WDM channels in routing components and in the receiver. However, the efforts to increase the spectral efficiency have led to sophisticated shaping of the optical spectrum. Techniques such as orthogonal frequency-division multiplexing and Nyquist WDM have demonstrated optical signal spectra that are very close to rectangular~\cite{schmogrow_2011_oe}. Using a DSP filter, the channel crosstalk can then be very small in the receiver. Optical routing components, such as reconfigurable optical add-drop multiplexers, are more difficult to realize as the filter function is implemented in optical hardware, but the lack of WDM channel spectral overlap reduces the crosstalk also here. Signal degradation due to, \textit{e.g.}, polarization-mode dispersion is neglected as it is compensated for by the receiver equalizer. Thus, we here choose to focus on the nonlinear effects generated as the interplay between the Kerr nonlinearity and the chromatic dispersion, known as the nonlinear interference (NLI) within the GN model~\cite{poggiolini_2012_jlt}.

\subsection{Modeling Aim}
The aim of the modeling effort is to find an approximate quantitative model for the NLI for a large number of connections between network transceivers. For each connection there is a route, \textit{i.e.}, a set of fiber links that connect the transmitting and the receiving nodes via a number of intermediate nodes. Each link can transmit $M$ WDM channels and for each channel $m = 1,\ldots,M$, the center frequency $f_m$, the bandwidth $\Delta f_m$, and the power $P_m$ are chosen.  This is summarized as the set of channel parameters $\conn_m = \{f_m, \Delta f_m, P_m\}$. For a given link, the WDM channels can then be written as the set $\wdm = \{\conn_1,\ldots,\conn_M\}$. The physical parameters of a link are the power attenuation $\alpha(z)$, the group-velocity dispersion $\beta_2(z)$, and the nonlinearity $\gamma(z)$.  Here $z \in [0, L]$ denotes the distance from the beginning of the first fiber, where $L$ is the total length of the link (possibly including several fiber spans). The link parameters are summarized in the set $\link(z) = \{\alpha(z), \beta_2(z), \gamma(z), L\}$.

Assume that there are a number of planned connections. To make sure that each transmission can succeed, the corresponding connection must satisfy the SNR requirement for the selected modulation format at the receiving node.  The purpose of the proposed model is to calculate the SNR for a number of simultaneous connections in a network, given the network topology and the parameter sets $\wdm$ and $\link$ for every link in the network. To the best of our knowledge, no such model exits in the literature.

The total noise variance is the sum of all noise sources.
%Transmitter noise is added at the transmitting node with a given PSD.
We view amplifier noise as being added after the power gain in each optical amplifier with a variance related to the power gain and the amplifier noise figure~\cite{agrawal_2010_focs}.  The NLI is generated during the propagation through the fibers, but within the GN model, we view it as being added after the gain in the amplifier, \textit{i.e.}, at the same location and at the same power level as the amplifier noise.

%There is the possibility to use repeaters to improve the signal quality or frequency conversion to change the center frequency. These components can both be included in the general framework outlined above. For example, a repeater can be modeled as an extra node and by demanding that the channel center frequency is never changed throughout the route, wavelength conversion is excluded.

\subsection{Modeling Approach: The GN Model}
We assume there is no periodic (hardware) compensation for chromatic dispersion, which seems to be a likely scenario for the future, and that no DSP compensation of nonlinear effects is carried out. Under these assumptions, the GN model is the current state of the art and in fact also seems to be the only approach that can be simplified to a sufficiently low-complexity model. While there are still open questions~\cite{dar_2013_arxiv} about the range of validity of the signal model used in the GN model~\cite[Eq.~(3)]{poggiolini_2011_ptl2}, the agreement with simulations of the full model equations has been shown to be very good~\cite{carena_2012_jlt}.  We discuss some known facts about the validity of the GN model in Section~\ref{sec_discussion}. Starting from the GN model, assumptions and approximations needed to obtain a network model are introduced in the following.

\section{Network Model Derivation}
\label{sec_model}

\subsection{General NLI Expression}
The derivation is started from the GN model as stated in~\cite{johannisson_2013_jlt}. As coherent systems typically transmit and receive the two polarization-multiplexed channels simultaneously, these are viewed as a unit. Assuming that the signal power spectral density (PSD) is equal in both polarizations, \textit{i.e.}, $\Gx(f) = \Gy(f) \equiv \Gsig(f)$, we use~\cite[Eqs.~(26) and (27)]{johannisson_2013_jlt} to obtain the NLI PSD
\begin{align}
\label{eq_total_psd}
\Gxptilde(f) &= 3 \iint |C(\fa, \fb, \fa + \fb - f)|^2
\nonumber \\
& \qquad \times \Gsig(\fa) \Gsig(\fb) \Gsig(\fa + \fb - f) \, d\fa d\fb,
\end{align}
where
\begin{align}
\label{eq_channel_func}
&C(\fa, \fb, \fa + \fb - f) = \nonumber \\
&\quad \int_{0}^{\Ltot} \gamma(z) p(z) e^{-i 4 \pi^2 (\fa - f) (\fb -
    f) \Ba(z)} \, dz.
\end{align}
Here, $\Gxptilde$ is the NLI PSD\footnote{The subscript ``1'' in the NLI PSD indicates that the result comes from a first-order perturbation analysis.} in the $x$ polarization but $\Gx(f) = \Gy(f)$ implies that $\Gxptilde(f) = \Gyptilde(f) \equiv \Gnli(f)$. The definitions for the normalized power evolution function, $p(z)$, and the accumulated dispersion, $\Ba(z)$, are given in~\cite[Section~II-B]{johannisson_2013_jlt}. Notice that the channel parameter $P_m$ is the power corresponding to $\Gx$, \textit{i.e.}, the power \emph{per polarization}.  The third-order dispersion has been neglected as it has very small impact on the NLI~\cite{johannisson_2013_jlt}. Finding the NLI PSD at a given frequency involves evaluating three nested integrals. The numerical complexity of this is high, which motivates our search for analytical simplifications of this expression.

\subsection{NLI Accumulation}
\label{sec_nli_acc}
The GN model was originally derived for a single link where all WDM channels propagate together from the transmitter to the receiver, but in the network model we need the capability of WDM channel switching/routing. The GN model builds on a certain signal model~\cite[Eq.~(3)]{poggiolini_2011_ptl2}, where a fundamental assumption is that the launched signal can be written on (or, at least, quickly approaches) this form. Then, during propagation, the changes of the launched signal are accounted for by a linearized propagation equation. As the signal leaves one fiber span and enters the next, the phase relation between different frequency components is conserved and this leads to a coherent accumulation of the complex amplitude of the signal perturbation corresponding to the NLI. In a network, channels can be added or dropped, which is incompatible with the original GN model.

Poggiolini \textit{et al.}\ have investigated the consequences from assuming that the NLI generated in different fiber spans can be added \emph{incoherently}, \textit{i.e.}, that the NLI PSDs are summed instead of the corresponding complex amplitudes~\cite{poggiolini_2012_jlt}. This assumption results in a significant analytical (and numerical) simplification, and numerical simulations have shown that the difference compared with the exact result is small. From the discussion above it is clear that in a network, the situation is neither fully coherent nor fully incoherent, but we will use the assumption that the NLI accumulates incoherently. It is then possible to calculate the NLI PSD generated in each channel in each fiber span separately and then sum these contributions along the entire route. Unfortunately, this assumption seems to slightly underestimate the NLI~\cite[Section~IX]{poggiolini_2012_jlt}.

For a given link, the NLI PSD $\Gnli(f)$ is a function of $\link$ and $\wdm$. The dependence on the link is static since the hardware does not change, but $\Gnli$ must be easy to evaluate as a function of $\wdm$.  The assumption of incoherent NLI accumulation implies that we can split each link $\link$ into $N$ spans $\link_n$, $n = 1,\ldots, N$, calculate the NLI independently for each span, and obtain the total NLI for a link from the NLI contributions from each span. In the special case that all spans are identical, it is obtained that $\Gnli = N \Gnli(f)$~\cite[Eq.~(18)]{poggiolini_2012_jlt}.

We assume that the loss of each fiber span is exactly compensated for by an EDFA placed at the end of the span.\footnote{While Raman amplification can be described within the GN model, it leads to the introduction of new system parameters and complicates the analytical expressions. It is outside the scope of the work presented here.} The received power is thus equal to the transmitted power $P$. Furthermore, since the link parameters $\link_n$ are assumed to be constant within a fiber span, the $z$ dependence is dropped from now on. Without loss of generality, we will therefore calculate the NLI generated in a single fiber span in Section~\ref{sec:span} and then in Section~\ref{sec:network-model} extend these results to the network layer.

\section{NLI for a Single Fiber Span}
\label{sec:span}

In this section, the NLI is derived for all channels in a single fiber span. The inputs are the channel parameters $\wdm$, and the link parameters $\link_n$. The output is the NLI PSD $\Gnli(f_m)$ for $m = 1,\ldots,M$. Although this case has been discussed in the literature, we study it in detail for two reasons. First, no expression seems to be published that gives the result for all channels in a flexible-grid WDM system, which is necessary for a complete network model. For example, Poggiolini \textit{et al.}\ mainly present results for the center WDM channel, see, \textit{e.g.}, \cite[Eqs.~(13)--(15)]{poggiolini_2012_jlt}. The second reason is that we find it useful to summarize all calculations to make this document self-contained.

%\subsection{Exact Integration}
As seen from~(\ref{eq_total_psd}), the signal PSD for each individual WDM channel must be known.  The exact shape depends on the transmitter, but as mentioned, much effort is currently spent on making the optical signal spectra close to rectangular~\cite{schmogrow_2011_oe}. Since this shape can be used with very small guard bands and also leads to minimal channel crosstalk, it is likely that signals in future systems will approximate this shape. Thus, we assume that each channel spectrum is rectangular, which implies that the channel PSD is completely specified by $\conn_m$ as $\Gsig(f_m) = P_m/\Delta f_m$. The channel PSDs may have different center frequencies, bandwidths, and powers, as exemplified in Fig.~\ref{fig_approx_int_signal_psd}.

From \eqref{eq_total_psd} and \eqref{eq_channel_func}, the NLI is
\begin{align}
\label{eq_nli}
\Gnli(f) &= \iint
%_{-\infty}^{\infty}
\frac{3 \gamma^2 \Gsig(\fa) \Gsig(\fb) \Gsig(\fa + \fb -
f)}{\alpha^2 + 16 \pi^4
  \beta_2^2 (\fa - f)^2 (\fb - f)^2}  \, d\fa d\fb,
\end{align}
where $1 - e^{-(\alpha + i 4 \pi^2 \beta_2 (\fa - f) (\fb - f)) L} \approx 1$ was used. This approximation is accurate for a fiber loss of 7~dB or more~\cite[Section~XI-A]{poggiolini_2012_jlt}. %Despite integrating over the entire $\fa \times \fb$ plane, the integral is finite due to the integrand's quadratic decay with $\fa$ and $\fb$.
The NLI is frequency-dependent and the channel $\conn_m$ is affected by the amount of NLI that passes the corresponding receiver filter.  Due to the assumption about the signal PSD shape, the matched filter has a rectangular frequency response. As it is difficult to analytically account for the variation of $\Gnli(f)$ within a channel, we assume that the NLI variance for channel $\conn_m$ can be approximated as $\Gnli(f_m) \, \Delta f_m$, \textit{i.e.}, that it can be based on the value at the center frequency.  This is a good approximation, as the NLI varies slowly within a given channel see, \textit{e.g.}, \cite[Fig.~5]{poggiolini_2012_jlt} and \cite[Fig.~1]{johannisson_2013_jlt}.  Furthermore, this assumption is conservative in the sense that it typically leads to a slight overestimation of the NLI.

\subsection{Approximate Integration}
The next step is to find an approximate value for $\Gnli(f_m)$ for channel $m$. This is done by generalizing the approach previously used for the center channel~\cite{poggiolini_2012_jlt, savory_2013_ptl}. As seen in (\ref{eq_nli}), the integration is over the entire $\fa \times \fb$ plane, but the integrand is nonzero only within distinct regions determined by the product of the PSDs. To exemplify, we consider channel $m = 2$ in the set of WDM channels in Fig.~\ref{fig_approx_int_signal_psd}. The PSD product in \eqref{eq_nli} is a piecewise constant function that is illustrated in the $\fa \times \fb$ plane in Fig.~\ref{fig_approx_int_ch_2_fwm_eff_and_domain} by the dark gray regions. The vertical, horizontal, and diagonal regions illustrate $\Gsig(\fa)$, $\Gsig(\fb)$, and $\Gsig(\fa + \fb - f_m)$, respectively. The product of the PSDs is nonzero only where three regions overlap. It is seen that this corresponds to polygons of different shapes, making exact integration difficult. The integrand weight function
\begin{align}
\label{eq_weight}
w_m(\fa, \fb) = \frac{3 \gamma^2}{\alpha^2 + 16 \pi^4 \beta_2^2 (\fa -
  f_m)^2 (\fb - f_m)^2}
\end{align}
is illustrated by the colored contours
%\footnote{The center of the weight function (as well as the second diagonal region) is centered over the second channel due to our choice to illustrate for this particular case.}
and in order to discuss its properties we introduce
\begin{align}
\eta_m &= \frac{w(f_m + \Delta f_m/2, f_m + \Delta f_m/2)}{w(f_m, f_m)} \nonumber \\
&= \left( 1 + \frac{\pi^4 \beta_2^2 \Delta f_m^4}{\alpha^2} \right)^{-1}.
\end{align}
In this way, $\eta_m$ is the fraction of the weight function evaluated at the edge of the $m$th channel spectrum relative to its channel center frequency. Using a dispersion parameter $D = 16$~ps/(nm$\,$km) and attenuation $\alpha = 0.2$~dB/km, we find $\eta_m \approx 0.84$ for a 10~GHz channel and $\eta_m \approx 0.078$ for a 28~GHz channel. When the values of $|\fa - f_m|$ and $|\fb - f_m|$ are increased, $w_m$ decreases rapidly.

%% p.const.c      = 2.99792458e8;             % Speed of light [m/s]
%% p.const.lambda = 1.55e-6;                  % Wavelength [m]
%% p.conv.D_to_beta2    = -p.const.lambda^2/2/pi/p.const.c*1e-6; % * D [ps/nm/km]
%% p.conv.att_DB_to_att = log(10)/10/1e3; % * alpha_dB [dB/km]
%% p.smf.beta2 = 16*p.conv.D_to_beta2;         % Group-velocity dispersion [s^2/m]
%% alpha_smf   = 0.20;                           % Attenuation [dB/km]
%% p.smf.alpha = alpha_smf*p.conv.att_DB_to_att; % Attenuation [1/m]
%% p.smf.gamma = 1.2e-3;                         % Kerr nonlinearity [1/W/m]
%% df = 28e9/2;
%% epsilon = 1/(1 + 16*pi^4*p.smf.beta2^2*df^4/p.smf.alpha^2)

We proceed by assuming that (i) only the polygons containing $\fa = f_m$ or $\fb = f_m$ need to be included and (ii) each polygon can be approximated by a rectangle of minimal area that contains the polygon. The first of these assumptions is equivalent to including \emph{self-channel interference} (SCI), represented by the single polygon that surrounds $(\fa, \fb) = (f_m, f_m)$, and \emph{cross-channel interference} (XCI). A similar approximation was discussed and illustrated in~\cite[Fig.~3]{poggiolini_2012_jlt} for the center channel when the channels have equal bandwidths and spacing. This approximation typically leads to a very small error, since $w_m$ is negligible elsewhere. The second assumption leads to an overestimation of the NLI. For example, in the SCI case, the polygon area is extended by a factor $4/3$, but the effect from this is reduced since $w_m$ is highest in the center. As shown above, this has some impact on the result for narrow-band channels.  Improvements of this approximation are possible, for example following~\cite{savory_2013_ptl}, but here the simple and conservative assumption above is used.

% Please don't move figures, it messes up the diff in version control
\newcommand{\mysize}{\footnotesize}
\begin{figure}[t]
\begin{center}
\includegraphics[width=0.99\linewidth, trim=8mm 4mm 0mm 0mm]{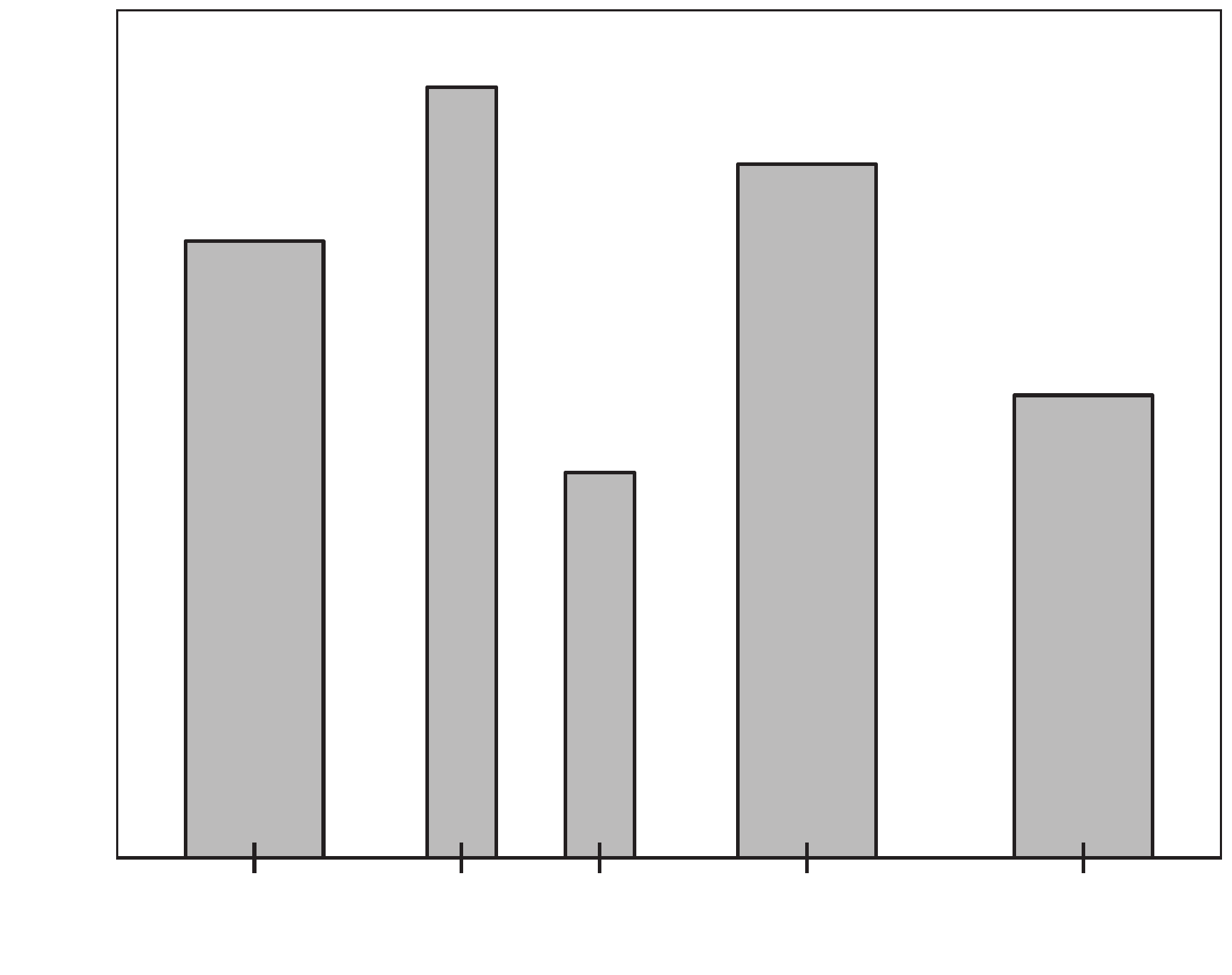}%
\overlay{-74.2mm}{3mm}{\mysize$f_1$}%
\overlay{-58.8mm}{3mm}{\mysize$f_2$}%
\overlay{-48.4mm}{3mm}{\mysize$f_3$}%
\overlay{-33mm}{3mm}{\mysize$f_4$}%
\overlay{-12.4mm}{3mm}{\mysize$f_5$}%
\overlay{-47.4mm}{0mm}{\mysize frequency $f$}%
\overlay{-88mm}{20mm}{\rotatebox{90}{\mysize power spectral density $\Gsig(f)$}}%
\caption{The PSD of a flexible-grid example WDM system, representing a set of $M = 5$ channels with different center frequencies, bandwidths, and powers.}
\label{fig_approx_int_signal_psd}
\end{center}
\end{figure}

% Please don't move figures, it messes up the diff in version control
\begin{figure}[t]
\begin{center}
\includegraphics[width=0.99\linewidth, trim=9mm 6mm 0mm 0mm]{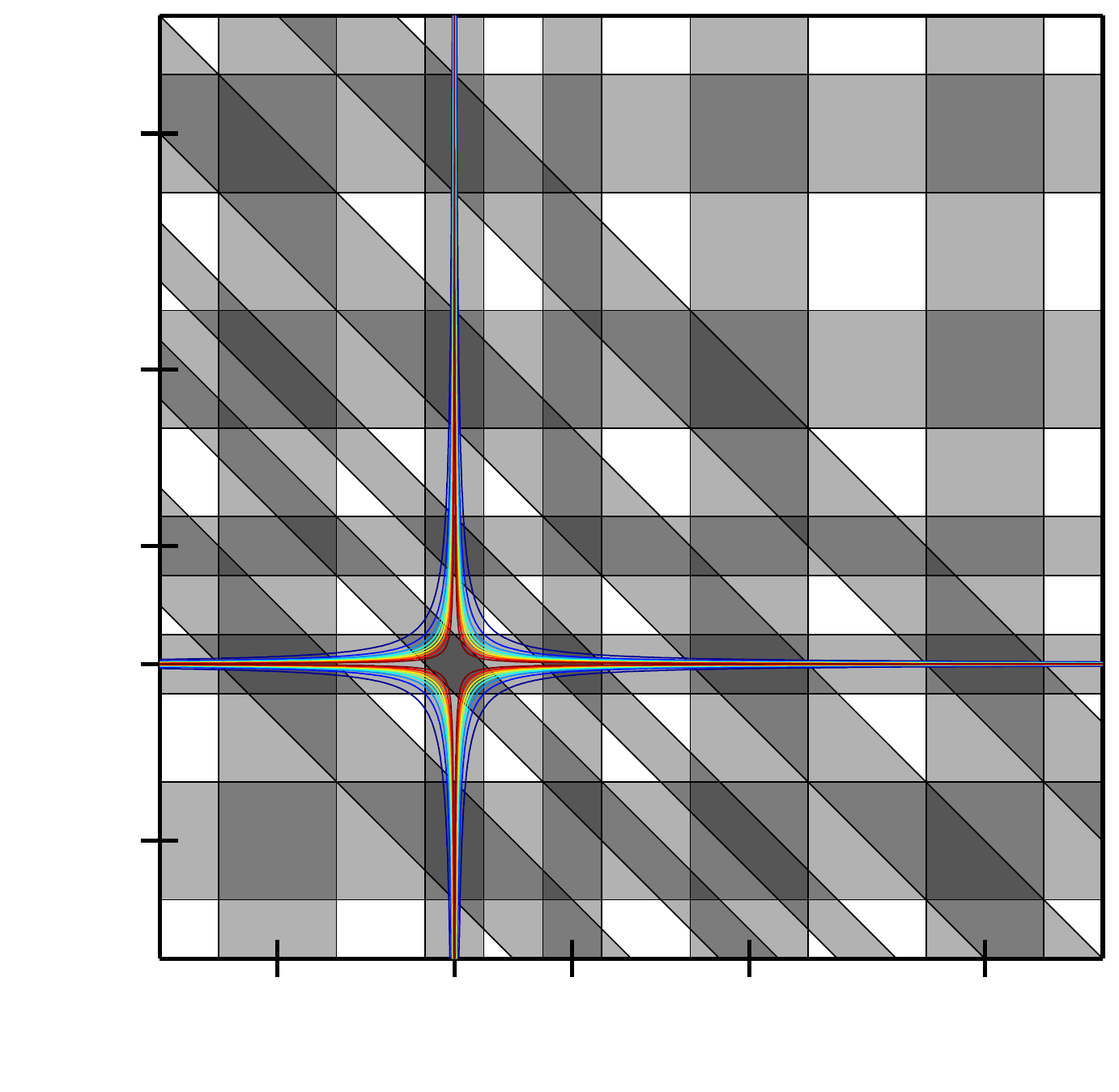}% lbrt
\overlay{-71.8mm}{3mm}{\mysize$f_1$}%
\overlay{-57.0mm}{3mm}{\mysize$f_2$}%
\overlay{-47.0mm}{3mm}{\mysize$f_3$}%
\overlay{-32.2mm}{3mm}{\mysize$f_4$}%
\overlay{-12.6mm}{3mm}{\mysize$f_5$}%
\overlay{-39mm}{0mm}{\mysize$\nu$}%
\overlay{-85mm}{15.8mm}{\rotatebox{90}{\mysize$f_1$}}%15.8
\overlay{-85mm}{30.6mm}{\rotatebox{90}{\mysize$f_2$}}%
\overlay{-85mm}{40.4mm}{\rotatebox{90}{\mysize$f_3$}}%
\overlay{-85mm}{55.2mm}{\rotatebox{90}{\mysize$f_4$}}%
\overlay{-85mm}{75.0mm}{\rotatebox{90}{\mysize$f_5$}}%
\overlay{-88mm}{46mm}{\rotatebox{90}{\mysize $\nu'$}}%
\caption{In this visual representation of (\ref{eq_nli}), the vertical, horizontal, and diagonal shaded regions correspond to $\Gsig(\fa)$, $\Gsig(\fb)$, and $\Gsig(\fa + \fb - f)$, respectively, where we choose $f = f_2$. This illustrates the qualitative behavior of the numerator in \eqref{eq_nli} for the example WDM channels in Fig.~\ref{fig_approx_int_signal_psd}.  Only the intersections of these regions contribute to the integral. The contours show the weight function from (\ref{eq_weight}), supporting the notion of considering only the regions containing $\fa = f_m$ or $\fb = f_m$.}
\label{fig_approx_int_ch_2_fwm_eff_and_domain}
\end{center}
\end{figure}

Inspection of (\ref{eq_nli}) shows that the result is unchanged if $\fa$ and $\fb$ are substituted for each other, which in Fig.~\ref{fig_approx_int_ch_2_fwm_eff_and_domain} corresponds to mirroring the entire integration domain in the line $\fa = \fb$. Thus, evaluating the NLI for channel $m$, the total NLI can be written as
\begin{align}
\Gnli(f_m) &= \Gsci + \sum_{\substack{m' = 1 \\ m' \neq m}}^{M} \Gxci,
\end{align}
where
\begin{align}
\Gsci &= \iint_{\Omega_{mm}} w_m(\fa, \fb) \Gsig^3(f_m) \,
d\fa d\fb, \\
\Gxci &= 2 \iint_{\Omega_{mm'}} w_m(\fa, \fb) \Gsig(f_m)
\Gsig^2(f_{m'}) \, d\fa d\fb,
\end{align}
and the integration domain $\Omega_{m m'}$ is the rectangle $\fa \in [f_m - \Delta f_m/2, f_m + \Delta f_m/2]$, $\fb \in [f_{m'} - \Delta f_{m'}/2, f_{m'} + \Delta f_{m'}/2]$.
By introducing
\begin{align}
\xi = \frac{4 \pi^2 |\beta_2|}{\alpha}
\end{align}
and
\begin{align}
\label{eq:F2}
F^2_{mm'} = \iint_{\Omega_{mm'}} \frac{1}{1 + \xi^2 (\fa - f_m)^2 (\fb
  - f_m)^2} d\fa d\fb,
\end{align}
it follows that
\begin{align}
\label{eq_G1}
\Gnli(f_m) &= \frac{3 \gamma^2}{\alpha^2} F^2_{m m} \Gsig^3(f_m) \nonumber \\
& \quad +
\sum_{\substack{m' = 1 \\ m' \neq m} }^{M}
\frac{6 \gamma^2}{\alpha^2} F^2_{m m'} \Gsig(f_m) \Gsig^2(f_{m'}).
\end{align}

Unfortunately, the exact result for $F^2_{m m'}$ has to be expressed in terms of the dilog function, which is defined in terms of a power series as $\dilog(z) = \sum_{n = 1}^{\infty} z^n/n^2$. For notational convenience, we introduce
\begin{align}
x_1 &= \frac{\Delta f_m}{2} \left( f_m - f_{m'} +
  \frac{\Delta f_{m'}}{2} \right) \xi, \\
x_2 &= \frac{\Delta f_m}{2} \left(f_{m'} - f_m + \frac{\Delta
  f_{m'}}{2} \right) \xi
\end{align}
to write \eqref{eq:F2} as
\begin{align}
F^2_{m m'} &= \frac{i}{\xi} [
\dilog ( -i x_1 ) - \dilog ( i x_1 ) \nonumber \\
& \qquad + \dilog (-i x_2 ) - \dilog (i x_2 )
].
\end{align}
From this expression, it is not obvious that $F^2_{m m'}$ is a real quantity. This can be made clear by rewriting it in terms of another special function as
\begin{align}
\label{eq_F2}
F^2_{m m'} &= \frac{1}{2\xi} \left[ x_1 \Phi(-x_1^2, 2, 1/2) +
  x_2 \Phi(-x_2^2, 2, 1/2) \right].
\end{align}
where $\Phi$ is the Lerch transcendent with the definition $\Phi(z, s, a) = \sum_{n = 0}^{\infty} z^n/(n + a)^s$, \cite[Section~9.55]{gradshteyn_1980_tables}.  Using (\ref{eq_F2}) together with (\ref{eq_G1}), the model is complete.

%% p.const.c      = 2.99792458e8;             % Speed of light [m/s]
%% p.const.lambda = 1.55e-6;                  % Wavelength [m]
%% p.conv.D_to_beta2    = -p.const.lambda^2/2/pi/p.const.c*1e-6; % * D [ps/nm/km]
%% p.conv.att_DB_to_att = log(10)/10/1e3; % * alpha_dB [dB/km]
%% p.smf.beta2 = 16*p.conv.D_to_beta2;         % Group-velocity dispersion [s^2/m]
%% alpha_smf   = 0.2;                           % Attenuation [dB/km]
%% p.smf.alpha = alpha_smf*p.conv.att_DB_to_att; % Attenuation [1/m]

%% epsilon = 0.5;
%% (p.smf.alpha^2/16/pi^4/p.smf.beta2^2*(1/epsilon - 1))^0.25

%% epsilon = 0.1;
%% (p.smf.alpha^2/16/pi^4/p.smf.beta2^2*(1/epsilon - 1))^0.25

%% df = 10e9/2;
%% 1/(1 + 16*pi^4*p.smf.beta2^2*df^4/p.smf.alpha^2)

%% df = 28e9/2;
%% 1/(1 + 16*pi^4*p.smf.beta2^2*df^4/p.smf.alpha^2)

\subsection{Further Simplification}
The result above has been presented in terms of the special function $\Phi$. In order to obtain a more intuitive expression, avoiding any special functions, we can approximate one step further. This can be done with an asymptotic expansion of the dilog function. Using the result from Appendix~\ref{sec_asymp}, (\ref{eq_G1}) can be written as
%% \begin{align}
%% &\Gnli(f_m) = \frac{3 \gamma^2 G_0(f_m)}{2 \pi \alpha |
%%   \beta_2|} \Bigg\{ G_0^2(f_m) \ln \left| \frac{\Delta f_m}{2}
%% %
%% \frac{2 \pi^2 \beta_2 \Delta f_m}{\alpha} \right| \nonumber \\
%% %
%% & + \sum_{m' = 1 \atop m \neq
%%   m'}^{M}
%% %
%% G_0^2(f_{m'}) \Bigg[ \ln \left| \left( f_{m m'} + \frac{\Delta
%%   f_{m'}}{2} \right) \frac{2 \pi^2 \beta_2 \Delta f_m}{\alpha}
%% \right| \nonumber \\
%% %
%% & \qquad - \ln \left| \left( f_{m m'} - \frac{\Delta f_{m'}}{2} \right) \frac{2
%%   \pi^2 \beta_2 \Delta f_m}{\alpha} \right| \Bigg] \Bigg\},
%% \end{align}
\begin{align}
\label{eq_G1simpler}
\Gnli(f_m) &= \frac{3 \gamma^2 \Gsig(f_m)}{2 \pi \alpha |
  \beta_2|} \Bigg( \Gsig^2(f_m) \ln \left|
\frac{\pi^2 \beta_2 (\Delta f_m)^2}{\alpha} \right| \nonumber \\
&\quad + \sum_{\substack{m' = 1 \\ m' \neq m}}^{M}
\Gsig^2(f_{m'}) \ln \left| \frac{f_{m m'} + \Delta f_{m'}/2}{f_{m
      m'} - \Delta f_{m'}/2} \right| \Bigg) ,
\end{align}
where we introduced $f_{m m'} \equiv |f_m - f_{m'}|$.
%% It is of course possible to rewrite the XCI term as a single logarithm of a fraction, but written in this way it is easy to see how the NLI from each WDM channel is obtained by evaluating the logarithm function at values corresponding to the spectrum edges.

\section{Summary of the Network Model}
\label{sec:network-model}

We are now ready to summarize the model and extend it to a network of multiple spans and links. The description in this section is intended to be self-contained and can be implemented without studying the details of the derivation in previous sections. It is evaluated in two stages; first to evaluate the channel quality of every \emph{link} in the network, and second, to evaluate the quality of every \emph{connection}.

The first stage utilizes the following inputs for a given link, denoted by $l$, in the network:
\begin{itemize}
\item The link parameters $\link_n$ of every fiber span $n=1,\ldots,N$. These are assumed to be the same for all channels.
\item The (constant) PSD $\Gase$ of the amplifier noise. It is equal to the sum of the noise PSDs added by the amplifier in each span, which are given by the amplifier gains and noise figures~\cite{agrawal_2010_focs}.
\item The channel parameters $\conn_m$ for $m=1,\ldots,M$. These are assumed to be constant through all fiber spans.
\end{itemize}
Given these quantities, the model is evaluated as follows:
\begin{enumerate}
\item Calculate $\Gsig(f_m) = P_m/\Delta f_m$ for $m=1,\ldots,M$.
\item For $n=1,\ldots,N$ and $m=1,\ldots,M$, use $\link_n$ to evaluate \eqref{eq_G1}, where $F^2_{mm'}$ is given by \eqref{eq_F2}. Denote the PSD $\Gnli(f_m)$ returned by \eqref{eq_G1} by $\Gnli_{mn}$.
\item For $m=1,\ldots,M$, calculate the SNR as $\mathit{SNR}_{lm} = P_m/\sigma_m^2$, where
\begin{align}
\sigma_m^2 = \Delta f_m \left(\Gase +\sum_{n=1}^N \Gnli_{mn}\right)
\end{align}
represents the total linear and nonlinear noise contributions.
\end{enumerate}
This is repeated for every link $l$ in the network. Alternatively, \eqref{eq_G1} and \eqref{eq_F2} may be replaced by \eqref{eq_G1simpler}, which is simpler but less accurate.

In the second stage, the SNR of an arbitrary connection in the network is calculated. The network is all-optical and assuming no format conversion, the bandwidth $\Delta f_m$ is constant for all links in the connection. The center frequency $f_m$ may change in nodes due to (ideal) wavelength conversion. The power $P_m$ is constant, as we have assumed the gains to balance the losses. However, our model holds also if $P_m$ changes between the links in a route, if the noise added in this process is negligible compared to the total $\Gase$.

The input to the second stage can be represented as follows.
\begin{itemize}
\item The route of the connection, represented as $K$ link assignments $l=l_1,\ldots,l_K$ and channel assignments $m=m_1,\ldots,m_K$.
\item The link and channel SNRs $\mathit{SNR}_{lm}$ obtained from the first stage above.
\end{itemize}
Based on this input, the SNR of the connection under consideration is finally obtained as
\begin{align}
\mathit{SNR} = \left(\sum_{k=1}^K\frac{1}{\mathit{SNR}_{l_k m_k}} \right)^{-1}
.\end{align}

%Let us select one connection from a transmitting to a receiving node. The SNR at the receiver is then
%\begin{align}
%\text{SNR} = \frac{P}{\sigma^2_\text{ASE} + \sigma^2_\text{NLI}},
%\end{align}
%where $P$ is the received signal power (which is equal to the transmitted power) and $\sigma^2_\text{ASE}$ and $\sigma^2_\text{NLI}$ are the variances of the amplifier noise and the NLI, respectively. We have $\sigma^2_\text{ASE} = G_\text{ASE} \, \Delta f$, where the PSD of the amplifier noise, $G_\text{ASE} = \sum_k G_{\text{ASE}, k}$, is the sum of the noise PSDs added by each individual amplifier in the connection route. These, in turn, are given by the amplifier gains and noise figures~\cite{agrawal_2010_focs}. For the NLI, we have $\sigma^2_\text{NLI} = G_\text{NLI} \, \Delta f$, where $G_\text{NLI} = \sum_l G_{1, l}$ and the summation includes each fiber span in each link in the route of the connection, which, in turn, is calculated using (\ref{eq_F2}), (\ref{eq_G1}), and the local expressions for $\link$ and $\wdm$.

\section{Discussion and Conclusions}\label{sec_discussion}

The described model is simple enough to allow optimization studies of optical networks operating in the nonlinear regime. As has been made clear in the derivation, a number of assumptions and approximations have been introduced. It is not possible to quantify the impact of these without performing a detailed numerical or experimental study, which is outside the scope of this work and we will instead qualitatively discuss the model accuracy.

First, the model obviously relies on the GN model and already (\ref{eq_total_psd}) and (\ref{eq_channel_func}) are approximate expressions. A fundamental assumption is that the dispersive effects are strong. This is difficult to formulate strictly but the GN model should not be used in systems with periodic dispersion compensation. Furthermore, the signal bandwidth should be sufficiently large and this means that single-channel transmission is less accurately modeled. While these assumptions are likely to be compatible with future optical networks, there are some further questions about the model validity. For example, it is a surprising fact that the model has no dependence on the choice of modulation format. The discussion about the validity of the GN model is ongoing and we expect that these questions will eventually be resolved.

The impact of the assumption about incoherent noise accumulation is difficult to quantify. It is known that this does introduce errors and the special case expression from Section~\ref{sec_nli_acc} is more accurately written as $\Gnli(f, \link, \wdm) = N^{1 + \epsilon} \Gnli(f, \link_1, \wdm)$, where $\epsilon \in [0, 1]$ depends on both the signal and the system~\cite[Section~IX]{poggiolini_2012_jlt}. There seems to be no exact analytical expression for $\epsilon$ available and this approach would also require all spans in the links to be identical. We see no obvious way to improve this approximation but it should be remembered that the WDM channel switching will reduce the error from this assumption.

Finally, approximations were introduced in the integration of \eqref{eq_nli}.  As seen, the value of $w_m$ is quickly reduced as the frequency separation is increased. Thus we expect the choice to include only SCI and XCI to lead to very small error, as long as not too narrow channel bandwidths are considered, but this is, as discussed, an inherent assumption of the GN model. For the same reason, the error introduced by approximating the integration polygons by rectangles is small.

We do not expect the proposed model to be the last word in the development of nonlinear fiber-optic network models, but rather a starting point. It is, to our knowledge, the first model that can predict the signal quality independently for a number of heterogeneous channels in a flexible-grid WDM system. Such a model is essential for efficient, if not optimal, resource management in elastic optical networks, which is an interesting and emerging area for future research.

\appendices
\section{Asymptotic Expansion}
\label{sec_asymp}
For $|z| \gg 1$, the dilog function has the asymptotic expansion
\begin{align}
\dilog(z) = \sum_{k = 0}^{\infty} (-1)^k (1 - 2^{1 - 2 k})(2 \pi)^{2 k}
\frac{B_{2 k}}{(2 k)!} \frac{[\ln(-z)]^{2 - 2 k}}{\Gamma(3 - 2 k)},
\end{align}
where $B_{2 k}$ are the Bernoulli numbers. As the $\Gamma(z)$ function is not defined for negative integers,
there are only two terms in the expansion, giving asymptotically
\begin{align}
\dilog(z) = -\frac{\pi^2}{6} - \frac{1}{2} \ln^2(-z).
\end{align}
However, assuming $x \in \mathbb{R}$, we have
\begin{align}
\ln (i x) &= \ln |x| + i \arg (i x) = \ln |x| + s_x i \frac{\pi}{2}, \\
\ln (-i x) &= \ln |x| + i \arg (-i x) = \ln |x| - s_x i \frac{\pi}{2},
\end{align}
where $s_x$ is the sign of $x$.  We get
\begin{align}
\dilog(i x) %&= -\frac{\pi^2}{6} - \frac{1}{2} \ln^2(-i x)  \nonumber \\
&=
-\frac{\pi^2}{6} - \frac{1}{2} \left( \ln |x| - s_x i \frac{\pi}{2}
\right)^2 \nonumber \\
%
%&= -\frac{\pi^2}{6} - \frac{1}{2} \left( \ln^2 |x| - 2 s_x i
%\frac{\pi}{2} \ln |x| -\frac{\pi^2}{4}\right) \nonumber \\
%
%&= - \frac{1}{2} \ln^2 |x| + s_x i \frac{\pi}{2} \ln |x| +
%\frac{\pi^2}{8} - \frac{\pi^2}{6} \nonumber \\
%
&= - \frac{1}{2} \ln^2 |x| + s_x i \frac{\pi}{2} \ln |x| -
\frac{\pi^2}{24}
\end{align}
and
\begin{align}
\dilog(-i x) &= -\frac{1}{2} \ln^2 |x| - s_x i \frac{\pi}{2} \ln |x| -
\frac{\pi^2}{24},
\end{align}
which gives
\begin{align}
F^2_{mm'} %&= \frac{i}{\xi} [ \dilog ( -i x_1 ) - \dilog ( i x_1
%  ) + \dilog (-i x_2 ) - \dilog (i x_2 ) ] \nonumber \\
%
&= \frac{\pi}{\xi} (s_{x_1} \ln |x_1| + s_{x_2} \ln |x_2|).
\end{align}

\ifCLASSOPTIONcaptionsoff
  \newpage
\fi



%\bibliographystyle{IEEEtran}
%\bibliography{IEEEabrv,../references}
%\begin{thebibliography}{1}
%\end{thebibliography}

\begin{IEEEbiographynophoto}{Pontus Johannisson} received his Ph.D.\ degree from Chalmers University of Technology, Gothenburg, Sweden, in 2006. His thesis was focused on nonlinear intrachannel signal impairments in optical fiber communications systems.

In 2006, he joined the research institute IMEGO in Gothenburg, Sweden, where he worked with digital signal processing for inertial navigation with MEMS-based accelerometers and gyroscopes. In 2009, he joined the Photonics Laboratory, Chalmers University of Technology, where he currently holds a position as Assistant Professor. A significant part of his time is spent working on cross-disciplinary topics within the Fiber-Optic Communications Research Center (FORCE) at Chalmers. His research interests include, \textit{e.g.}, nonlinear effects in optical fibers and digital signal processing in coherent optical receivers.
\end{IEEEbiographynophoto}

%\newpage

\begin{IEEEbiographynophoto}{Erik Agrell}
(M’99--SM’02) received the Ph.D.\ degree in information theory in 1997 from Chalmers University of Technology, Sweden.

From 1997 to 1999, he was a Postdoctoral Researcher with the University of California, San Diego and the University of Illinois at Urbana-Champaign. In 1999, he joined the faculty of Chalmers University of Technology, first as an Associate Professor and since 2009 as a Professor in Communication Systems. In 2010, he cofounded the Fiber-Optic Communications Research Center (FORCE) at Chalmers, where he leads the signals and systems research area. His research interests belong to the fields of information theory, coding theory, and digital communications, and his favorite applications are found in optical communications.

Prof.\ Agrell served as Publications Editor for the IEEE Transactions on Information Theory from 1999 to 2002 and is an Associate Editor for the IEEE Transactions on Communications since 2012. He is a recipient of the 1990 John Ericsson Medal, the 2009 ITW Best Poster Award, the 2011 GlobeCom Best Paper Award, the 2013 CTW Best Poster Award, and the 2013 Chalmers Supervisor of the Year Award.
\end{IEEEbiographynophoto}

\end{document}